\documentstyle[12pt,twoside]{article}
\def\greaterthansquiggle{\raise.3ex\hbox{$>$\kern-.75em\lower1ex\hbox{$\sim$}}}
\def\lessthansquiggle{\raise.3ex\hbox{$<$\kern-.75em\lower1ex\hbox{$\sim$}}}
\newcommand{\beq}{\begin{equation}}
\newcommand{\eeq}{\end{equation}}
\newcommand{\beqa}{\begin{eqnarray}}
\newcommand{\eeqa}{\end{eqnarray}}
\newcommand{\beqan}{\begin{eqnarray*}}
\newcommand{\eeqan}{\end{eqnarray*}}
\newcommand{\ba}{\begin{array}}
\newcommand{\ea}{\end{array}}
\newcommand{\no}{\nonumber}

\newcommand{\ve}{\varepsilon}
\newcommand{\vp}{\varphi}

\newcommand{\wt}{\widetilde}

\newcommand{\B}{{\cal B}}

\newcommand{\cL}{{\cal L}}

\newcommand{\R}{{\cal R}}

\newcommand{\st}{\stackrel}

\def\nz{\ifmmode {I\hskip -3pt N} \else {\hbox {$I\hskip -3pt N$}}\fi}
\def\zz{\ifmmode {Z\hskip -4.8pt Z} \else
       {\hbox {$Z\hskip -4.8pt Z$}}\fi}
\def\qz{\ifmmode {Q\hskip -5.0pt\vrule height6.0pt depth 0pt
       \hskip 6pt} \else {\hbox
       {$Q\hskip -5.0pt\vrule height6.0pt depth 0pt\hskip 6pt$}}\fi}
\def\rz{\ifmmode {I\hskip -3pt R} \else {\hbox {$I\hskip -3pt R$}}\fi}
\def\cz{\ifmmode {C\hskip -4.8pt\vrule height5.8pt\hskip 6.3pt} \else
       {\hbox {$C\hskip -4.8pt\vrule height5.8pt\hskip 6.3pt$}}\fi}

\normalsize
\begin{document}
\bibliographystyle{plain}
\begin{titlepage}
\begin{flushright}
UWThPh-1994-50\\
\today
\end{flushright}
\vspace{2cm}
\begin{center}
{\Large \bf
The Momentum Constraints of General Relativity \\[5pt]
and Spatial Conformal Isometries}\\[50pt]
R. Beig*  \\
Institut f\"ur Theoretische Physik \\
Universit\"at Wien \\
Boltzmanngasse 5, A--1090 Wien, Austria\\
Fax: ++43-1-317-22-20, E-mail: BEIG@PAP.UNIVIE.AC.AT \\[5pt]
and \\[5pt]
N. \'O Murchadha \\
Physics Department \\
University College Cork \\
Cork, Ireland
\vfill
{\bf Abstract} \\
\end{center}

Transverse--tracefree (TT--) tensors on $({\bf R}^3,g_{ab})$, with
$g_{ab}$ an asymptotically flat metric of fast decay at infinity, are
studied. When the source tensor from which these TT tensors are
constructed has fast fall--off at infinity, TT tensors allow a
multipole--type expansion. When $g_{ab}$ has no conformal Killing
vectors (CKV's) it is proven that any finite but otherwise arbitrary
set of moments can be realized by a suitable TT tensor. When CKV's
exist there are obstructions --- certain (combinations of) moments
have to vanish --- which we study.
\begin{center}
MSC numbers: 83C05, 83C40
\end{center}
\vfill
\noindent *) Supported by Fonds zur F\"orderung der wissenschaftlichen
Forschung in \"Osterreich, Project No. P9376--PHY.
\end{titlepage}

\section{Introduction}
\renewcommand{\theequation}{\arabic{section}.\arabic{equation}}
\setcounter{equation}{0}

In this paper we consider transverse--tracefree (TT--) tensors on
${\bf R}^3$ with an asymptotically flat metric $g_{ab}$, i.e. tensors $P_{ab}$
satisfying
\beq
D^a P_{ab}  = 0 , \qquad \mbox{trace } P = 0 \quad \mbox{on }
({\bf R}^3,g_{ab}),
\eeq
where $D$ is the covariant derivative associated with $g$. The interest in
this problem
comes first of all from (vacuum) general relativity, where Equ. (1.1)
is the momentum constraint for an initial data set $({\bf R}^3,g_{ab},P_{ab})$
\beq
D^a (P_{ab} - g_{ab} \mbox{ trace } P) = 0
\eeq
in the maximal (i.e. trace $P = 0$) case. As is well--known, Equ. (1.2)
is just the expression of the invariance of the theory under diffeomorphisms
of three space. Thus our study of Equ. (1.1) is relevant to a much larger
class of theories than Einstein's.

In the standard conformal approach to solving
the constraints, Equ. (1.1) is not solved on the physical metric $g_{ab}$,
but a conformally related metric $g'_{ab}$ having faster decay at infinity
than $g_{ab}$. One is here using the fact that $P_{ab}$ being $TT$
is invariant under $g'_{ab} = \omega^2 g_{ab}$, $P'_{ab} = \omega^{-1}
P_{ab}$, $\omega > 0$. We call $g'_{ab}$, $P'_{ab}$ again $g_{ab}$,
$P_{ab}$. Our assumptions on $g_{ab}$ are that $g_{ab}$ is smooth and,
in standard coordinates $x^a$ on ${\bf R}^3$, satisfies
\beq
g_{ab} - \delta_{ab} =
O^\infty \left( \frac{1}{r^{K-1+\ve}}\right),
\qquad 0 < \ve < 1
\eeq
for some $K = 1,2,\ldots$, where $r = (x^a x^b \delta_{ab})^{1/2}$ and
$F = O^\infty(f(r))$ means that $F = O(|f(r)|)$,
$\partial F = O(|f'(r)|)$, $\partial \partial F = 0(|f''(r)|)$, a.s.o.
In addition we require a condition of conformal smoothness for $g_{ab}$,
as follows: there are functions
$$
f^a(x) = O^\infty \left( \frac{1}{r^{K-2+\ve}}\right),
$$
such that, with $\bar x^a = x^a + f^a(x)$,
$\Omega^{-1} = \delta_{ab} x'{}^a x'{}^b$, the tensor field
$\wt g_{ab} = \Omega^2 g_{ab}$ admits a smooth extension in coordinates
$\wt x^a = x^a/\Omega$ to $\wt x^a = 0$.
For example, these assumptions will be valid for all $K$ when
$g_{ab}$ equals the flat metric outside a compact subset of ${\bf R}^3$.
For $P_{ab}$ we require that
\beq
P_{ab} = O^\infty \left( \frac{1}{r^2} \right).
\eeq
We shall impose one more condition on $P_{ab}$ which arises as follows.
Any smooth, trace--free tensor $Q_{ab}$ satisfying (1.4) can be written
as (see Chaljub--Simon [3])
\beq
Q_{ab} = P_{ab} + (LW)_{ab} ,
\eeq
where
\beq
(LW)_{ab} := D_a W_b + D_b W_a - \frac{2}{3} g_{ab} D_c W^c ,
\eeq
i.e. the conformal Killing operator associated with the vector field
$W^a$ satisfying
\beq
W_a = O^\infty \left( \frac{1}{r} \right)
\eeq
and $P_{ab}$ being $TT$. Thus
\beq
D^b (LW)_{ab} = \Delta W_a + \frac{1}{3} D_a (D^b W_b) + \R_a{}^b W_b
= D^b Q_{ab},
\eeq
where $\R_a{}^b$ is the Ricci tensor of $g_{ab}$.
Given $Q_{ab}$, $W_a$ and whence $P_{ab}$ is unique. Thus the decomposition
(1.5) can be used to find $TT$--tensors and, clearly, all $TT$--tensors
arise this way (just take $Q_{ab} = P_{ab}$, $W_a = 0$!).
We call $Q_{ab}$ a ``source tensor'' for $P_{ab}$. It now seems natural
to restrict $P_{ab}$ further by imposing asymptotic conditions on the
source tensor from which it arises. We assume
\beq
Q_{ab} = O^\infty \left( \frac{1}{r^{1+K+\ve}} \right)
\eeq
where $\ve$ and $K$ are the same numbers as the ones appearing in (1.3).
As $K$ increases we shall obtain more detailed information on the
multipole behaviour of $W_a$, and whence $P_{ab}$, near infinity. We
will then ask and answer the question whether, given arbitrary values for the
relevant multipole moments, whose number depends on $K$, a source tensor
$Q_{ab}$ satisfying (1.9) can be found, yielding a $P_{ab}$ having
precisely these moments.
Since the map sending $Q_{ab}$ to
$P_{ab}$ is many--to--one: $Q_{ab}$ and $Q_{ab} + (Ls)_{ab}$ for any
$s_a$ satisfying $s_a = O^\infty(1/(r^{1+K+\ve}))$ give the same
$TT$--tensor, one wonders what, if anything, condition (1.9) means in
terms of $P_{ab}$. The answer is that (1.3,5,7,9) imply
\beq
\Delta D_{[a} P_{b]c} = O^\infty \left( \frac{1}{r^{4+K+\ve}}\right),
\eeq
and we state without proof that the converse also holds.
(The appearance of third derivatives in Equ. (1.10) is no accident.
If we viewed $P_{ab}$ as a linearization of the metric $g_{ab}$,
the left hand side in (1.10) is essentially the linearization of
the Cotton tensor, whose vanishing is equivalent to conformal
flatness of a metric. The connection between TT--tensors and the
linearized Cotton tensor will be further studied in forthcoming
work by one of us (R.B.).)
Conditions such as (1.9), while natural from the viewpoint of the
decomposition (1.5), do not have an obvious physical interpretation. As a
model one could look at the Gau\ss\ constraint of electrodynamics
\beq
D^a E_a = 0 \qquad \mbox{on } ({\bf R}^3, \delta_{ab})
\eeq
and the ansatz
\beq
E_a = q_a + D_a \phi
\eeq
with $q_a$ of fast fall--off, say of compact support. Thus
$D_{[a} E_{b]}$ has compact support. If the same assumption is
made for the magnetic fields $B_a$ one would find that $(E_a,B_a)$
are data for an electromagnetic field which is stationary in the
domain of dependence of a neighbourhood of infinity. Presumably,
in some approximate sense, a similar interpretation could be given
for (1.9), when supplemented by some additional conditions on the
metric $g_{ab}$ (see e.g. Reula [10]).

\section{The asymptotic expansion}
\renewcommand{\theequation}{\arabic{section}.\arabic{equation}}
\setcounter{equation}{0}

Let, first, $g_{ab}$ be the flat metric $\delta_{ab}$ on ${\bf R}^3$
and consider the elliptic equation
\beq
\partial^2 W_a + \frac{1}{3} \partial_a (\partial^b W_b) = j_a ,
\eeq
where $j_a$ is smooth and $j_a = O^\infty(1/(r^{2+K+\ve}))$,
$K = 1,2,\ldots$. The unique
solution $W_a$ to (2.1) going to zero at infinity is given by
\beq
W_a(x) = - \frac{1}{4 \pi} \int_{{\bf R}^3} F_{ab} (x - x')
j^b(x') d^3 x',
\eeq
where
\beq
F_{ab}(x) = \frac{1}{8} \left( 7 \frac{\delta_{ab}}{r} +
\frac{x_a x_b}{r^3} \right).
\eeq
It is straightforward to see from (2.1,2,3) that $W_a(x)$ admits an
expansion
\beq
\st{K}{W}_a(x) = \sum_{k=1}^K \frac{\st{k}{\omega}_a(n)}{r^k}
+ O^\infty \left( \frac{1}{r^{K+\ve}} \right) ,
\eeq
where $\st{k}{\omega}_a$ are smooth on $S^2$ ($n^a =: x^a/r$).
Instead of computing the vectors $\st{k}{\omega}_a$ in terms of
$j_a$ directly from (2.2,3) we prefer an apparently more roundabout
but actually
more efficient way, as follows. The vectors $\st{k}{\omega}_a$ can be
decomposed into parts orthogonal and tangential to $S^2$, i.e.
\beq
\st{k}{\omega}_a = n_a \st{k}{\sigma} + \st{k}{\mu}_a.
\eeq
The tangential parts $\st{k}{\mu}_a$, in turn, can be expanded as
\beq
\st{k}{\mu}_a = \nabla_a \st{k}{\vp} + \ve_a{}^b \nabla_b
\st{k}{\psi},
\eeq
where $\nabla$ is the derivative on $S^2$ and $\ve_{ab}$ the volume
element on $S^2$. The scalars $\st{k}{\sigma}$, $\st{k}{\vp}$ and
$\st{k}{\psi}$ can now be expanded in terms of spherical harmonics,
e.g.
\beq
\st{k}{\vp} = \sum_{\ell =0}^\infty \st{k}{m}_{a_1 \ldots a_\ell}
n^{a_1} \ldots n^{a_\ell}
\eeq
where $\st{k}{m}_{a_1 \ldots a_\ell}$ are symmetric, trace--free tensors.
Putting all this back into $\st{k}{\omega}_a$ we see that
\beq
\st{k}{\omega}_a = \sum_{\ell = 0}^\infty
(n_a \st{k}{M}_{b_1 \ldots b_\ell} n^{b_1} \ldots b^{b_\ell} +
\st{k}{O}_{ab_1 \ldots b_\ell} n^{b_1} \ldots n^{b_\ell} +
\ve_a{}^{bc} n_b \st{k}{L}_{c b_1 \ldots b_\ell} n^{b_1} \ldots
n^{b_\ell}),
\eeq
$$ 1 \leq k \leq K $$
where all of $\st{k}{M}$, $\st{k}{O}$, $\st{k}{L}$ are symmetric and
trace--free. We now insert (2.4,8) into (2.1). It follows that the
first term in Equ. (2.4) has to satisfy (2.1) with $j_a = 0$. This
results in a coupling between the number $k$ in (2.8) and the
$\ell$--values which can give a contribution. More precisely, we find
after a straightforward computation that
$\st{k}{L}_{a b_1 \ldots b_\ell} = 0$ except for $k = \ell + 2$. We also find
that $\st{k}{M}_{b_1 \ldots b_\ell}$ and $\st{k}{O}_{a b_1 \ldots b_\ell}$
both = 0 except for $k = \ell$ in which case
\beq
(8 - k) \st{k}M_{b_1 \ldots b_k} - (2k -1) \st{k}O_{b_1 \ldots b_k} = 0, \qquad
k \geq 1
\eeq
or $k = \ell + 2$ in which case
\beq
(k - 2) \st{k}{M}_{b_1 \ldots b_{k-2}} + (2k - 3) \st{k}{O}_{b_1 \ldots
b_{k-2}} = 0 .
\eeq
Thus $\st{K}{W_a}$ can be written as a sum of three terms plus a
remainder, i.e.
\beq
\st{K}{W_a} = \st{(1)}{W_a} + \st{(2)}{W_a} + \st{(3)}{W_a} +
O^\infty \left( \frac{1}{r^{K+\ve}} \right) ,
\eeq
where
\beqa
\st{(1,K)}{W_a} &=& \sum_{k=2}^K \frac{\ve_a{}^{bc} n_b
\st{k}{L}_{c b_1 \ldots b_{k-2}} n^{b_1} \ldots n^{b_{k-2}} }{r^k} \\
\st{(2,K)}{W_a} &= & \sum_{k=1}^K \frac{(2k-1) n_a
\st{k}{M}_{b_1 \ldots b_k} n^{b_1} \ldots n^{b_k} + (8-k)
\st{k}{M}_{a b_1 \ldots b_{k-1}} n^{b_1} \ldots n^{b_{k-1}} }{r^k} \\
\st{(3,K)}{W_a} &= & \sum_{k=2}^K \frac{(2k-3) n_a
\st{k}{N}_{b_1 \ldots b_{k-2}} n^{b_1} \ldots n^{b_{k-2}} - (k-2)
\st{k}{N}_{a b_1 \ldots b_{k-3}} n^{b_1} \ldots n^{b_{k-3}} }{r^k} .\no \\
\eeqa
We now observe that the estimates (2.11 -- 14) remain valid, when
the l.h. side of (2.1) is replaced by (1.8), where the flat metric is
replaced by one satisfying (1.3) and all of (2.12,13,14) is
understood with respect to the flat background metric at infinity.
We have thus proven
\paragraph{Theorem 1:} Let $W_a$ be a solution of
\beq
\Delta W_a + \frac{1}{3} D_a (D^b W_b) + \R_a{}^b W_b = j_a ,
\eeq
$$
g_{ab} - \delta_{ab} = O^\infty \left( \frac{1}{r^{K-1+\ve}}\right)
\qquad \mbox{and} \qquad
j_a = O^\infty \left( \frac{1}{r^{2+K+\ve}} \right), \qquad
K = 1, \ldots ,
$$
with $W_a$ going to zero at infinity. Then there is a string of
``multipole moments'' $L$, $M$ and $N$, such that Equ.'s (2.11 -- 14)
are valid.

At this stage it is important to remark that the source $j_a$ in Equ.
(2.1) does not necessarily come from a $Q_{ab}$, s. th. $j_a =
D^b Q_{ab}$ satisfying (1.9) for the respective $K$. As an example
consider $Q_{ab}$ of the form ($F_{abc}$ are symmetric, trace--free
constants)
\beq
\bar Q_{ab} = \frac{2 F_{abc} n^c - 6 n_{(a} F_{b)cd} n^c n^d +
3(\delta_{ab} - n_a n_b) F_{cde} n^c n^d n^e}{r^2}
\eeq
for $r > R > 0$, and extended smoothly as a tracefree tensor to all
of ${\bf R}^3$.
$\bar Q_{ab}$ so chosen satisfies the flat--space equation
$\partial^a \bar Q_{ab} = 0$ for $r > R$. Thus
$j_a = D^b \bar Q_{ab}$ is $O^\infty (1/(r^{2+K+\ve}))$ and
$P_{ab} = \bar Q_{ab} + (LW)_{ab}$ satisfies the momentum constraints
together with (1.4). But it is not of the form
$(L \st{k}{W})_{ab} + O^\infty (1/(r^{1+K+\ve}))$ for $K = 1,2,\ldots$,
since $\bar Q_{ab}$ is only $O^\infty(1/r^2)$. Thus neither $P_{ab}$, nor
$\bar Q_{ab}$, satisfy (1.10) for any $K$, and this can of
course also be checked by direct computation.

There is a second and more fundamental way in which (2.1) can fail to
solve our original problem. This can occur when $j_a$ is such that we
have difficulty finding a trace--free $Q_{ab}$ for
which $j_a = D^b Q_{ab}$. This can occur when $(M,g_{ab})$ has conformal
isometries, i.e. conformal Killing vectors (CKV's) $\xi^a$:
\beq
(L\xi)_{ab} = D_a \xi_b + D_b \xi_a - \frac{2}{3} g_{ab} D_c \xi^c = 0.
\eeq
Let $\xi^a$ be any such vector field. Then
\beq
-\int_{{\bf R}^3} \xi^a D^b Q_{ab} dV + \oint_{r = \infty} \xi^a Q_{ab}
dS^b = \int_{{\bf R}^3} (D^a \xi^b) Q_{ab} dV  .
\eeq
Thus, from Eq. (2.17), the left hand side of (2.18) is zero. If we were to do
this analysis on a compact manifold without boundary, the surface term in
equ.(2.18) would not appear and we get the immediate restriction that $j_a$
must be $L^2$--orthogonal to $\xi^a$. In the asymptotically flat case this
restriction gets softened to the requirement that if $j_a$ is not orthogonal
to $\xi^a$ then the falloff of $Q_{ab}$ must be slow enough that the surface
integral in (2.18) does not vanish.

Equ.(2.18) has a second use. If $Q_{ab}$ is source-free, i.e., is a
TT--tensor, we see that the surface integral in Equ.(2.18) must vanish,
irrespective of the decay rate of $Q_{ab}$. This will have further
consequences.

The possible
existence of CKV's will be important in our next goal, which is trying to find
$Q_{ab}$'s, for which the moments appearing in (2.11,12,13) assume arbitrary
values. To see this we write out the lowest two orders in this expansion, i.e.
for $K = 2$ \beqa
\st{(1,2)}{W_a} & = & \frac{\ve_a{}^{bc} n_b \st{2}{L}_c}{r^2} +
O^\infty \left( \frac{1}{r^{2 +\ve}} \right) \\
\st{(2,2)}{W_a} & = & \frac{n_a \st{1}{M}_b n^b + 7 \st{1}{M}_a}{r} +
\frac{3n_a \st{2}{M}_{bc} n^b n^c + 6 \st{2}{M}_{ab} n^b}{r^2} +
O^\infty \left( \frac{1}{r^{2 +\ve}} \right) \\
\st{(3,2)}{W_a} & =&  \frac{n_a \st{2}{N}}{r^2} +
O^\infty \left( \frac{1}{r^{2 +\ve}} \right) .
\eeqa
Let $\xi^a_T$ be an asymptotic translation, i.e. a vector field of the
form
\beq
\xi^a_T = \mu^a + O^\infty \left( \frac{1}{r^\ve} \right) ,
\eeq
where the $\mu^a$'s are constants. Then, using the decay of $Q_{ab}$,
\beq
\oint_{r=\infty} P_{ab} \xi^a_T dS^b = \oint_{r = \infty}
(L \st{(2,2)}{W})_{ab} \xi^a_T dS^b = - 32 \pi \st{1}{M}_a \mu^a .
\eeq
Thus $\st{1}{M}_a$ is essentially the (conserved) ADM 3--momentum.
(In order to compare with the standard definition one has to
check that the same value is obtained,
when one takes $P_{ab}$ in (2.23) to be the physical extrinsic
curvature $\bar P_{ab}$ related to $P_{ab}$ by $\bar P_{ab} = \phi^{-2}
P_{ab}$, where $\phi$ is the solution to the Lichnerowicz equation.)
If $\xi_T{}^a$ happens to be a CKV and if we have no source--current, the l.h.
side of (2.22) is zero, and we obtain the obstruction $\st{1}{M}_a \mu^a = 0$.

Suppose, next, that we have an asymptotic rotation vector $\xi_R{}^a$, i.e.
\beq
\xi_R{}^a = \ve^{abc} x_b \kappa_c + O^\infty
\left( \frac{1}{r^\ve} \right) .
\eeq
Then
\beq
\oint_{r=\infty} P_{ab} \xi_R{}^a dS^b = \int_{r = \infty}
(L \st{(1,2)}{W})_{ab} \xi_R{}^a dS^b = - 8 \pi \st{2}{L}_a \kappa^a.
\eeq
Thus $\st{2}{L}_a$ is essentially the conserved ADM 3--angular momentum.
When $\xi_R{}^a$ is a CKV and the matter is at rest, we have the obstruction
$\st{2}{L}_a \kappa^a = 0$.

We will show in the next section that the quantities
$\st{1}{M}_a$, $\st{2}{L}_a$, $\st{2}{N}$ and $\st{3}{N}_a$ appearing in
$\st{(3,3)}{W}_a$ are the only ones which can possibly not be
specified arbitrarily. The essential step will be a description of all
moments in terms of surface integrals like (2.23), which will however
not be expressible just in terms of $P_{ab}$, but will involve both
$P_{ab}$ and $W_a$.

\section{The $\lambda$--fields}
\renewcommand{\theequation}{\arabic{section}.\arabic{equation}}
\setcounter{equation}{0}

Define the following collections of vector fields
\beqa
\st{(1,k)}{_o \lambda}_a &=& \ve_a{}^{bc} x_b
\st{k}{\kappa}_{c b_1 \ldots b_{k-2}} x^{b_1} \ldots x^{b_{k-2}},
\qquad k \geq 2  \\
\st{(2,k)}{_o \lambda}_a &=& \st{k}{\mu}_{a b_1 \ldots b_{k-1}}
x^{b_1} \ldots x^{b_{k-1}} , \qquad k \geq 1  \\
\st{(3,k)}{_o \lambda}_a &=& 2(7k - 11) x_a
\st{k}{\nu}_{b_1 \ldots b_{k-2}} x^{b_1} \ldots x^{b_{k-2}} \no \\
&& \mbox{} - (k-2)(k+7) r^2 \st{k}{\nu}_{a b_1 \ldots b_{k-3}}
x^{b_1} \ldots x^{b_{k-3}}, \qquad k \geq 2
\eeqa
where, again, the flat background metric is used and all of $\kappa$,
$\mu$, $\nu$ are constants which are symmetric and trace--free with
respect to that metric. These fields have the following properties:
they are globally regular (although they blow up at infinity) and
they are annihilated by the flat space operator (2.1). Thus
\beq
D^b(L \st{(\alpha,k)}{_o\lambda})_{ab} = O^\infty \left( \frac{1}
{r^{2+\ve}} \right)
\eeq
for $\alpha = 1,2,3$ provided that $K \geq k$. Thus, using
[3], we can uniquely solve the equations
\beq
\Delta  \st{(\alpha,k)}{\delta\lambda_a} + \frac{1}{3} D_a D^b
\st{(\alpha,k)}{\delta\lambda_b} + \R_a{}^b
\st{(\alpha,k)}{\delta \lambda_b}
= - D^b(L \st{(\alpha,k)}{\lambda})_{ab}
\eeq
with $ \st{(\alpha,k)}{\delta\lambda_a} = O^\infty (1/r^\ve)$.
Calling $\st{(\alpha,k)}{\lambda_a} = \st{(\alpha,k)}{_o \lambda_a}
+ (\st{(\alpha,k)}{\delta\lambda})_a$, we obtain the

\paragraph{Theorem 2:} For any non--zero choice of symmetric, trace--free
constants in (3.1,2,3) there exist unique non--zero vector fields
$\st{(\alpha,k)}{\lambda_a}$ with
\beq
\st{(\alpha,k)}{\lambda_a} = \st{(\alpha,k)}{_o \lambda_a} +
O^\infty \left( \frac{1}{r^\ve} \right),
\eeq
satisfying
\beq
D^b (L \st{(\alpha,k)}{\lambda})_{ab} = 0.
\eeq
Particularly interesting in this list are the ``special'' fields
$\st{(2,1)}{\lambda}$, $\st{(1,2)}{\lambda}$, $\st{(3,2)}{\lambda}$,
$\st{(3,3)}{\lambda}$ which we call --- in this order --- asymptotic
translations, rotations, dilations and conformal boosts.

\paragraph{Lemma 1:} If the manifold has a CKV and if $K \geq 3$, it must
be a linear combination of $\st{(\alpha,k)}{\lambda_a}$.

\paragraph{Proof:} We know that CKV's cannot go to zero at infinity
(Christodoulou and \'O Murchadha [4]). The CKV, call it
$\xi$, since it satisfies $L\xi = 0$, must satisfy $D(L\xi) = 0$. A
decomposition such as used in the proof of Theorem 1 shows that the leading
part of $\xi$ must be a $\st{(\alpha,k)}{_o \lambda_a}$. If we now demand that
the first order condition  ($L\xi = 0$) be satisfied, we find that only the
``special'' fields listed above can survive. The Lemma follows.

An alternative way of saying this is:

\paragraph{Lemma 2:} Let us have a linear combination
$\st{K}{\lambda} = \sum_{k=1}^K c(\alpha,k) \st{(\alpha,k)}{\lambda}$, $k \leq
K
   $,
$c(\alpha,k) =$ const, which, in addition to (3.7), satisfies the
strong condition $(L \st{K}{\lambda})_{ab} = 0$, i.e. is a CKV. Then,
trivially, for $K = 1$ it is an asymptotic translation. For $K = 2$,
$c(2,2) = 0$ and, for $K \geq 3$, all further $c(\alpha,k)$, $k \leq K$,
vanish except $c(3,3)$. In other words, $\st{K}{\lambda}$ can only be
a linear combination of the special vector fields.

The proof of this Lemma is a straightforward computation. Clearly,
when $g_{ab}$ is conformally flat, the special $\lambda$--fields are
all CKV's. When $g_{ab}$ is not conformally flat, almost the opposite
is true. We cannot have either the translation or the
dilation CKV's.  We can have (at most) only one rotation CKV and up to
three conformal boosts. Namely, we have

\paragraph{Theorem 3:} Let $g_{ab}$ not be conformally flat. Then none
of the $\st{(2,1)}{\lambda}$'s are CKV's. For $K \geq 2$, there is at most one
linear combination of $\st{(2,1)}{\lambda}$, $\st{(1,2)}{\lambda}$,
$\st{(3,2)}{\lambda}$ which can be a CKV, and this has to satisfy
$\st{2}{\nu} = 0$ (which means $\st{(3,2)}{\lambda} \equiv 0$) and there exists
a vector
$d^a$, such that
$\ve^a{}_{bc} d^b \st{2}{\kappa}{}^c = \st{1}{\mu}{}^a$.
In other words this CKV has to be an
asymptotic rotation, possibly after a shift of origin.

\paragraph{Proof:} Let us assume that there exists a CKV which blows up
like $r$ at infinity. From our conformal smoothness assumption on $g_{ab}$ it
follows (Geroch [6]) that this CKV $\lambda^a$ extends to a smooth CKV $\wt
\lambda^a$ for some smooth metric $\wt g_{ab}$ on ${\bf R}^3 \cup \{ r =
\infty\} \cong S^3$. From the asymptotic condition we have that $\wt
\lambda^a$  vanishes at the point--at--infinity $\Lambda$, i.e. $\left. \wt
\lambda^a \right|_\Lambda = 0$. Furthermore
\beq
\wt \lambda^a = \wt r^2 \st{1}{\mu}{}^a - 2 \wt x^a(\wt x_b \st{1}{\mu}{}^b)
+ \ve^a{}_{bc} \wt x^b \st{2}{\kappa}{}^c + 6 \st{2}{\nu} \wt x^a +
O^\infty (\wt r^{2+\ve}) .
\eeq
Invariantly, we have that
\beqa
\left. \wt D_a \wt \lambda^a \right|_\Lambda &=& 18 \st{2}{\nu} \no \\
\left. (\wt D_a \wt \lambda_b) \right|_\Lambda &=& \ve_{ab}{}^c
\st{2}{\kappa}_c =: F_{ab} \\
\left. (\wt D_a \wt D_b \wt \lambda^b)\right|_\Lambda &=&
- 2 \st{1}{\mu}_a . \no
\eeqa
Now we recall the notion of an inessential (resp. essential) CKV.
A CKV is called inessential, if there
exists a metric $g'_{ab} = \omega^2 \wt g_{ab}$, $\omega > 0$, so that
it is a Killing vector w.r. to $g'_{ab}$. Otherwise the CKV is called
essential. Suppose the CKV $\wt \lambda^a$ was inessential. This would
imply that
\beq
\left. D'_a \wt \lambda^a \right|_\Lambda = 0, \qquad
\left. (D'_a D'_b \wt \lambda^b)\right|_\Lambda = 0
\eeq
for some suitable conformal metric $g'$.
But under conformal rescalings, using $\left. \wt \lambda^a \right|_\Lambda
= 0$,
\beq
\left. D'_a \wt \lambda^a \right|_\Lambda =
\left. \wt D_a \wt \lambda^a \right|_\Lambda
\eeq
and
\beq
\left. (D'_a D'_b \wt \lambda^b) \right|_\Lambda =
\left.(\wt D_a \wt D_b \wt \lambda^b) \right|_\Lambda + \left. 3 F_a{}^b
(\omega^{-1} \wt D_b \omega)\right|_\Lambda .
\eeq
Thus, if $\wt \lambda^a$ is inessential, we would have that $\st{2}{\nu}$
is zero and $\st{1}{\mu}$ is of the form
$\st{1}{\mu}{}^a  = \ve^a{}_{bc} d^b \st{2}{\kappa}{}^c$
for some vector $d^a$. The only alternative is that
$\wt \lambda^a$ is essential.
But it is shown in Appendix A that this is impossible except if
$(\wt M, \wt g_{ab})$ is conformally diffeomorphic to $S^3$ with the
standard metric. This also follows from a famous result of Obata [9],
and Appendix A goes some way towards giving an independent proof of the
full Obata theorem in 3 dimensions.

In order to show the ``at most one''--statement in Theorem 3, suppose
there was a second CKV $\wt \lambda^a$ vanishing at $\Lambda$. By
taking the commutator between the two, we obtain a third such CKV.
Now, using (the full force of) the Obata theorem, their action, when
$\wt g_{ab}$ is not conformal to the standard metric on $S^3$, would
again have to be inessential, i.e. isometric after a conformal
rescaling. Since $\Lambda$ is fixed, this would have to be an action
under $SO(3)$ with $S^2$ principal orbits and thus (Fischer [5]) a
standard spherical action on $S^3$ with all orbits $S^2$ except for
two fixed points. Consequently, $\wt g_{ab}$ would have the standard
rotational symmetry and thus be conformal to the standard metric.
This contradiction  ends the proof of Theorem 3.

We add the following remark: When $\st{(1,2)}{\lambda}$ (a rotation) is a CKV
and, in addition, satisfies
\beq
\cL_{\st{(1,2)}{\lambda}} P_{ab} = - (D^c \st{(1,2)}{\lambda}_c) P_{ab}
\eeq
it follows that for the physical initial--data set
$\bar P_{ab} = \phi^{-2} P_{ab}$, $\bar g_{ab} = \phi^4 g_{ab}$,
with $\phi$ being the Lichnerowicz conformal factor,
$\st{(1,2)}{\lambda}$ as an isometry, i.e.
$\cL_{\st{(1,2)}{\lambda}} \bar g_{ab} =
\cL_{\st{(1,2)}{\lambda}} \bar P_{ab} = 0$.
But (3.13) implies that $\st{1}{\kappa}_a$ is parallel to $\st{2}{L}_a$
in the center of energy,
whereas, from (2.25), we have that $\st{2}{L}_a \st{1}{\kappa}{}^a$ is
zero. Thus $\st{2}{L}_a$ vanishes. It is in fact a known result, although we
are not aware of a place in the literature where this is explicitly
stated, that an asymptotically flat, topologically trivial vacuum spacetime
with a $U(1)$--isometry has zero angular momentum in the centre of
energy.

\section{The product $\langle \lambda|W\rangle$}
\renewcommand{\theequation}{\arabic{section}.\arabic{equation}}
\setcounter{equation}{0}

We now use the $\lambda$--vector fields to obtain a useful description
of the moments of $W_a$ in terms of surface integrals. Consider the
following antisymmetric scalar product
\beq
\left\langle \st{(\alpha,\ell)}{\lambda} | \st{K}{W} \right\rangle :=
\oint_{r = \infty} \left[ \st{(\alpha,\ell)}{\lambda^a}
(L \st{K}{W})_{ab} - \st{K}{W}{}^a (L \st{(\alpha,\ell)}{\lambda})_{ab}
\right] dS^b
\eeq
for $\ell \leq K$. Using (2.14) and (3.7) we see that
\beq
\left\langle \st{(\alpha,\ell)}{\lambda} | \st{K}{W} \right\rangle =
\int_{{\bf R}^3} \st{(\alpha,\ell)}{\lambda^a} \; j_a\; dV.
\eeq
In particular, since $j_a = O^\infty(1/(r^{2+K+\ve}))$ and
$\st{(\alpha,\ell)}{\lambda} = O^\infty (r^{\ell-1})$, the surface
integrals in (4.1) converge. The remainder terms in (2.10) and (3.6) do not
contribute to these integrals so
that they can be evaluated explicitly in terms of the constants
entering $\st{(\alpha,K)}{W}$ and $\st{(\alpha,k)}{\lambda}$. This is
a somewhat tedious exercise. We need the following crucial facts. Any
integral of the form
\beq
I(A,B) = \int_{S^2} A_{a_1 \ldots a_k} n^{a_1} \ldots n^{a_k}
B_{b_1 \ldots b_\ell} n^{b_1} \ldots n^{b_\ell} d^2S
\eeq
is zero for $k \neq \ell$, by virtue of orthogonality of spherical
harmonics ($A$ and $B$ are symmetric and trace--free). For $k = \ell$,
(4.3) can be computed (Appendix B), to give
\beq
I(A,B) = 4 \pi \frac{2^\ell (\ell!)^2}{(2 \ell + 1)!} A \cdot B \;
\delta_{k \ell},
\eeq
where $A \cdot B := A_{a_1 \ldots a_\ell} B^{a_1 \ldots a_\ell}$.
It is furthermore easy to see that an integral of the form
\beq
J(A,B) = \int_{S^2} \ve^{abc} A_{a a_1 \ldots a_k} n^{a_1} \ldots n^{a_k}
B_{b b_1 \ldots b_\ell} n^{b_1} \ldots n^{b_\ell} n_c d^2 S
\eeq
is zero for all $(k,\ell)$. It follows from the last remark that
$\langle\st{(\alpha,\ell)}{\lambda}|\st{(\beta,K)}{W}\rangle$ is zero,
when one of $(\alpha,\beta)$ is equal to one and the other is not.
Using (4.4) we find for $\alpha = 2$, $\beta = 3$ and for $\alpha = 3$,
$\beta = 2$ that all terms which remain after using orthogonality of
spherical harmonics in fact cancel. Thus
\beq
\left\langle \st{(\alpha,\ell)}{\lambda}|\st{(\beta,K)}{W} \right\rangle
= \delta_{\alpha \beta} F(\alpha,\ell), \qquad \ell \leq K
\eeq
where for $F(\alpha,\ell)$ we finally obtain
\beqa
F(1,\ell) &=& - 4\pi \frac{2^{\ell -1}[(\ell - 2)!]^2}{(2\ell - 3)!}
\left( \st{\ell}{L} \cdot \st{\ell}{\kappa} \right), \qquad
\ell \geq 2 \\
F(2,\ell) &=& - 4\pi \frac{2^{\ell +2}[(\ell - 1)!]^2}{(2\ell - 2)!}
 \st{\ell}{M} \cdot \st{\ell}{\mu} , \qquad
\ell \geq 1 \\
F(3,\ell) &=& - 4\pi \frac{(\ell^3 + 2 \ell^2 - 29\ell + 54)2^{\ell-2}
[(\ell - 3)!]^2}{(2 \ell - 5)!}
 \st{\ell}{N} \cdot \st{\ell}{\nu} , \qquad \ell \geq 3 \\
F(3,2) &=& - 4 \pi \cdot 36 \st{2}{N} \; \st{2}{\nu} . \no
\eeqa
In particular, $F(\alpha,\ell)$ are all non--zero. Thus the product
$\langle \lambda|W \rangle$ gives rise to a pairing between the moments
contained in $\st{(\alpha,\ell)}{\lambda}$ and those contained in
$\st{(\beta,K)}{W}$. So we can take two sets of basis vectors for the
two sets of symmetric, trace--free tensors involved, which are dual with
respect to this pairing. The dimension of each set can be computed
e.g. from (3.1,2,3): any symmetric, trace--free tensor with $k$ indices
contributes $2k +1$ dimensions. This gives
$$
\sum_{k=2}^K [2(k-1) + 1] = K^2 - 1 \mbox{ for } \st{(1,k)}{\lambda},
$$
$$
\sum_{k=1}^K (2k + 1) = K^2 + 2K \mbox{ for } \st{(2,k)}{\lambda}
$$
and
$$
\sum_{k=2}^K [2(k-2) + 1] = K^2 - 2K + 1 \mbox{ for } \st{(3,k)}{\lambda},
$$
adding up to $3K^2$ dimensions. This is of course consistent with the
three linear momentum components at order $1/r$ and the nine
independent moments at order $1/r^2$ in Equ.'s (2.19,20,21).
We thus have $\lambda$--fields $\st{A}{\lambda_a}$, with
$1 \leq A \leq 3K^2$. The $3K^2$ moments $\st{A}{M}$ encoded by the
terms of order $1/r$ up to order $1/r^K$ in $W$, in the above basis,
are
\beq
\st{A}{M} = \langle  \st{A}{\lambda} |W \rangle ,
\eeq
provided $j$ in (2.14) is $O^\infty(1/(r^{2+K+\ve}))$.

We now ask the question. Given a set of moments $\st{A}{M}$: does there
exist a source $j_a$ having the required fall--off, so that the unique
$W$ solving Equ. (2.14) has exactly these moments? The answer is
affirmative, as the following consideration shows. Take, for $j_a$,
the linear combination
\beq
j_a = \frac{1}{(1 + r^2)^{(2+2K'+\ve)/2}} \sum_{A=1}^{3K^2}
\st{A}{c}\; \st{A}{\lambda_a}, \qquad K' \geq K, \qquad c^A =
\mbox{const.}
\eeq
This is clearly $O^\infty(1/(r^{2+K+\ve}))$. Inserting (4.11) into (4.10),
using (4.2), we are now left with the finite--dimensional linear equation
\beq
\st{A}{M} = \sum_{B=1}^{3K^2} D_{AB} \; c^B,
\eeq
with $D_{AB} = D_{(AB)}$ given  by
\beq
D_{AB} = \int_{{\bf R}^3} \frac{1}{(1 + r^2)^{(2+2K'+\ve)/2}}
\st{A}{\lambda_a} \;\st{B}{\lambda^b} \; dV .
\eeq
The matrix $D$ is clearly positive definite:
$\sum_{A,B} D_{AB} c^A c^B = 0$ would imply $\sum c^A \st{A}{\lambda_a}
\equiv 0$,
and this, by using (3.1,2,3) at increasing orders in $1/r$ and using
the orthogonality of spherical harmonics, can only happen when
$c^A = 0$. Thus Equ. (4.12) can be uniquely solved for $c^A$, given
arbitrary $M^A$.

 We now come to the question of existence of $Q_{ab}$,
so that $W_a$, solving (2.14) with $j_a = D^b Q_{ab}$ has arbitrary
moments up to some finite order. The answer is afforded by

\paragraph{Theorem 4:}
\begin{enumerate}
\item[a)] All moments other than the ``exceptional moments'' $\st{1}{M}_a$,
$\st{2}{L}_a$, $\st{2}{N}$ for $K \geq 2$, plus $\st{3}{N}_a$ appearing in
$\st{(3,3)}{W}_a$ for $K \geq 3$ can be prescribed by a suitable choice
of $Q_{ab}$ satisfying (1.9).
\item[b)] When $g_{ab}$ is conformally flat, the exceptional moments are
constrained to vanish.
\item[c)] Suppose $g_{ab}$ is not conformally flat and $K \geq 2$: when
there are no conformal isometries, all of $\st{1}{M}_a$, $\st{2}{L}_a$,
$\st{2}{N}$ and $\st{3}{N}_a$
can be prescribed. Otherwise, by Theorem 3, there is at most one CKV
which  has $\st{3}{\nu}_a = 0$. This
has $\st{2}{\nu} = 0$ and, possibly after a shift of coordinates,
$\st{1}{\mu}_a = 0$.
In these coordinates $\st{1}{M}_a$, $\st{2}{L}_a$ and $\st{2}{N}$ can
still be prescribed with the only condition
that $\st{2}{L}_a \st{1}{\kappa}{}^a = 0$.
\end{enumerate}

\paragraph{Proof:} Again we start from
\beq
\st{A}{M} = \langle \st{A}{\lambda}|W\rangle = \int_{{\bf R}^3}
\st{A}{\lambda}{
   }^a
\; j_a \; dV .
\eeq
But, since $j_a = D^b Q_{ab}$, one more integration by parts yields
\beq
\st{A}{M} = - \int_{{\bf R}^3} (L \st{A}{\lambda})^{ab} \; Q_{ab} \; dV .
\eeq
When $g_{ab}$ is conformally flat, using the $A$--values corresponding to
the special $\lambda$--fields, we find that all exceptional moments are
zero, which proves b). To prove a) and c), we make the ansatz
\beq
Q_{ab} = \frac{1}{(1 + r^2)^{(2K'+\ve)/2}} \sum_{A=1}^{3K^2} \st{A}{c}
(L \st{A}{\lambda})_{ab} , \qquad K' \geq K, \qquad c^A = \mbox{const}
\eeq
and try to solve
\beq
\st{A}{M} = \sum_{B=1}^{3K^2} E_{AB} \; c^B ,
\eeq
where $E_{AB} = E_{(AB)}$ is defined by
\beq
E_{AB} = \int_{{\bf R}^3} \frac{1}{(1 + r^2)^{(2K'+\ve)/2}} (L
\st{A}{\lambda})^
   {ab}
(L \st{B}{\lambda})_{ab} dV .
\eeq
Equ. (4.17) can be solved provided $\sum M^A f^A = 0$, where
$\sum E_{AB} f^B = 0$. But the latter condition, by (4.18), implies that
$\sum f^A \st{A}{\lambda_a}$ is a CKV.
Choosing, successively, for $\st{A}{\lambda}$ all possibilities except for the
special $\lambda$--fields and using the pairing (4.6--9) and Lemma 2, we
see that a) is true. Using Theorem 3 statement c) follows similarly.

\paragraph{Remark:} If we insist on prescribing a non--zero value
for $\st{3}{N}_a$, we have to
take into account the possibility of CKV's with $\st{3}{\nu}_a \neq 0$.
Such cases do in fact exist (see Beig, Husa [2]), and give rise to
more conditions on the exceptional moments.

We end this paper with a ``compact--support version'' of Theorem 4,
namely
\paragraph{Theorem 4':} Let $g_{ab}$ be a metric on ${\bf R}^3$ which is
flat outside a compact set. Then all statements on arbitrariness of
moments in Theorem 4 remain valid, when $Q_{ab}$ is constrained to have
compact support, rather than the fall--off of Equ. (1.9).

\paragraph{Proof:} The ansatz (4.16) is now replaced by
\beq
Q_{ab} = \rho \sum_{A=1}^{3K^2} \st{A}{c} (L \st{A}{\lambda})_{ab},
\eeq
where $\rho \in C_0^\infty ({\bf R}^3)$, $\rho \geq 0$ and $\rho > 0$
in a region $\B$ strictly containing the support of $g_{ab} -
\delta_{ab}$. Then the null space of $E_{AB}$ consists of vectors $f^A$
such that $\sum f^A \st{A}{\lambda}_a = \lambda_a$ satisfies
$(L \lambda)_{ab} = 0$ in $\B$.
But, outside $\B$, $\lambda_a$ satisfies the flat--space version of
$D^a(L\lambda)_{ab} = 0$. Taking one more divergence of this equation
we see that $D_a \lambda^a$ is harmonic and, inserting back, that
each component $\lambda_a$ is harmonic. Thus $\lambda_a$, whence
$(L \lambda)_{ab}$ is analytic outside $\B$. It follows that
$(L \lambda)_{ab} = 0$ everywhere, and $\lambda^a$ is a CKV. Now the
statements of Theorem~4 follow, again, from Lemma~2 and Theorem~3.

\appendix{\section*{Appendix A}}

\newcounter{zahler}
\renewcommand{\thesection}{\Alph{zahler}}
\renewcommand{\theequation}{\Alph{zahler}.\arabic{equation}}
\setcounter{zahler}{1}
\setcounter{equation}{0}

Let $\wt M$ be a connected 3 dimensional compact manifold without
boundary, with a smooth metric $\wt g_{ab}$. Let $\xi^a$ be a CKV
on $(\wt M, \wt g_{ab})$. For discussing whether $\xi^a$ can be
essential, we distinguish between two cases, based on the sign of
$\lambda_1(\wt g)$, the lowest eigenvalue of the conformal Laplacian
$L_{\wt g} = - \Delta_{\wt g} + \frac{1}{8} \R[\wt g]$, where $\R$ is
the scalar curvature of $\wt g$. The first case is, from our present
viewpoint, the unphysical case, since the Hamiltonian constraint
cannot be solved for maximal data, when the background metric is
conformally extendable to a metric $\wt g$ on the compactified manifold
$\wt M$ with $\lambda_1(\wt g) \leq 0$.

\paragraph{Theorem A.1:} Let $\lambda_1(\wt g) \leq 0$. Then $\xi^a$ is
inessential.

\paragraph{Proof:} Let $\bar g$ be a metric conformal to $\wt g$ with
$\R[\bar g] =$ const. This exists by the easier part of the solution to
the Yamabe problem (Trudinger [12]). The rest is an argument due to
Lichnerowicz [8]. By straightforward computation we find from
\beq
\wt D_a \wt \xi_b + \wt D_b \wt \xi_a = \frac{2}{3} \wt g_{ab}
\wt D_c \xi^c
\eeq
and $\bar \R \equiv \R_0$ = const, that
\beq
\left(\Delta_{\bar g} + \frac{\R_0}{2} \right) \bar D_a \xi^a = 0.
\eeq
Since $\R_0 \leq 0$, the maximum principle implies that
$\bar D_a \xi^a =$ const and $\R_0 = 0$. Integrating $\bar D_a \xi^a$
over $(\wt M,\bar g)$ gives zero, by the Gau\ss\ theorem. Thus
$\bar D_a \xi^a = 0$, and $\xi^a$ is a Killing vector of $\bar g_{ab}$.

\paragraph{Theorem A.2:} Let $\lambda_1(g) > 0$ and $\xi^a$ be a CKV
vanishing at $\Lambda \in \wt M$. Then, either
\begin{enumerate}
\item[a)] $(\wt M, \wt g)$ is conformally diffeomorphic to $S^3$ with
the standard metric $\st{0}{g}_{ab}$. Or
\item[b)] $-3\alpha := \left. \wt D_a \xi^a\right|_\Lambda = 0$ and
$- 6 c_a := \left. \wt D_a \wt D_b \xi^b \right|_\Lambda$ lies in the
image of the linear map $F_a{}^b$ with
$F_{ab} := \left. \wt D_{[a} \wt \xi_{b]} \right|_\Lambda$.
\end{enumerate}

\paragraph{Proof:} Since $\lambda_1(\wt g) > 0$, the operator $L_[\wt g]$
has a positive Green function (see [7]), which we take to be centered
at $\Lambda$, i.e.
\beq
L_{\wt g} G = 4 \pi \delta_\Lambda .
\eeq
$G$ has the asymptotic expansion [7]
\beq
G = \frac{1}{\|x\|} + \frac{m}{2} + O^\infty(\|x\|),
\eeq
where $\|x\|$ is geodesic distance from $\Lambda$ and $m$ is the ADM
mass of the asymptotically flat metric
\beq
g'_{ab} = G^b \wt g_{ab} \quad \mbox{on } M = \wt M \setminus \Lambda .
\eeq
By virtue of (A.3) the metric $g'_{ab}$ has vanishing scalar curvature,
i.e. $\R[g'] = 0$.
The expansion (A.4) can be improved: under conformal rescalings
$\bar g_{ab} = \omega^2 g_{ab}$, $\omega > 0$, $G$ changes according to
$\bar G = \left. \omega^{-1/2}\right|_\Lambda \omega^{-1/2} G$.
Now $\omega > 0$ can be found so that $\R_{ab}[\bar g]$ is zero at
$\Lambda$. Thus, in this conformal gauge,
$\bar g_{ab} = \delta_{ab} + O(\|x\|^3)$ in Riemannian normal
coordinates $x^a$ centered at $\Lambda$. Expanding $L_{\bar g}$
accordingly and using standard estimates for the flat--space Green
function
$\st{0}{G}(x,x') = (x - x',x - x')^{-1/2}$, where
$(x,y) = \delta_{ab} x^a y^b$, it follows that $(|x| = (x,x)^{1/2})$
\beq
G = \frac{1}{|x|} + \frac{m}{2} + (d,x) + O^\infty(|x|^2) , \qquad
d^a = \mbox{ const.}
\eeq
For $\xi^a$ we have the expansion
\beq
\xi^a = - \alpha x^a + F^a{}_b x^b + c^a x^2 - 2 x^a(c,x) +
O^\infty(|x|^4) .
\eeq
(Note that, in the notation of (3.9), $\alpha = - 6 \st{2}{\nu}$,
$c^a = \frac{1}{3} \st{2}{\mu}{}^a$.)
The uniqueness of the Green function implies (see Beig [1]) that
$\xi^a$ is a homothetic vector field for the metric $g'_{ab}$.
Equivalently,
\beq
\cL_\xi G + \frac{1}{6} (\wt D_a \xi^a) G = \gamma G, \qquad
\gamma = \mbox{const.}
\eeq
Evaluating the l.h. side of Equ. (A.8) using (A.6, 7), we find that
\beq
\cL_\xi G = \frac{\alpha}{|x|} + \frac{(c,x)}{|x|} - \alpha(d,x) +
F^a{}_b x^b d_a + O^\infty(|x|^2)
\eeq
\beq
\frac{1}{6} (\wt D_a \xi^a)G = - \frac{\alpha}{2 |x|} - \frac{\alpha m}{4}
- \frac{(c,x)}{|x|} - \frac{(\alpha d + mc,x)}{2} +
O^\infty(|x|^2).
\eeq
Comparing coefficients in (A.8), there results $\alpha = 2 \gamma$ at
order $-1$ in $|x|$ and $\alpha m = - 2 \gamma m$ at order $0$. Thus,
either
\begin{enumerate}
\item[a)] $m = 0$: Then, since $\R[g'] = 0$, the positive mass theorem
[11] applies and yields that $(M,g')$
is diffeomorphic to ${\bf R}^3$ with the standard metric, or
$(\wt M,\wt g)$ conformally diffeomorphic to the standard $S^3$. Or,
\item[b)] $m > 0$: Then $\alpha = \gamma = 0$. The order 1 in (A.8) now
gives
\beq
F^b{}_a d_b = \frac{m}{2} c_a .
\eeq
\end{enumerate}
This ends the proof of Theorem (A.2): Clearly, transforming to
asymptotically flat coordinates $x'{}^a = x^a/|x|^2$ we see that
$\alpha$ corresponds to the dilation part, $F_{ab}$ the rotation
part and $c^a$ the translation part of the CKV on ${\bf R}^3$
associated with $\xi$.

\appendix{\section*{Appendix B}}

\renewcommand{\thesection}{\Alph{zahler}}
\renewcommand{\theequation}{\Alph{zahler}.\arabic{equation}}
\setcounter{zahler}{2}
\setcounter{equation}{0}

To prove (4.4) it suffices to consider $A = B$.
\beq
I(B,B) = B^{a_1 \ldots a_k} B^{a_{k+1} \ldots a_{2k}}
\int_{S^2} n_{a_1} \ldots n_{a_{2k}} d^2 S.
\eeq
The integral in (B.1) is proportional to
$$
\delta_{(a_1 a_2} \ldots \delta_{a_{2k-1} a_{2k)}}.
$$
Using the formula $(a \in {\bf R}^3)$
\beq
\int_{S^2} (a,n)^k d^2 S = \frac{4 \pi |a|^2}{2k+1} ,
\eeq
the proportionality constant is found to $4\pi/(2k + 1)$. It remains
to evaluate
$$
\frac{4\pi}{2k+1} B^{a_1 \ldots a_k} B^{a_{k+1} \ldots a_{2k}}
\delta_{(a_1 a_2} \ldots \delta_{a_{2k-1} a_{2k)}}.
$$
Of the $(2k)!$ terms in this expression, due to the vanishing trace of
$B$, only those terms contribute for which, in each Kronecker delta,
there is one $i_\ell$ with $1 \leq \ell \leq k$ and one $i_m$ with
$k + 1 \leq m \leq 2k$, of which there are $2^k(k!)^2$. Thus
\beq
I(B,B) = \frac{4 \pi 2^k (k!)^2}{(2k+1)!} B_{a_1 \ldots a_k}
B^{a_1 \ldots a_k} .
\eeq

\section*{Acknowledgements}
We thank Piotr Chrusciel for pointing out Ref. 5 and useful comments on
the manuscript and Walter Simon and Helmuth Urbantke for discussions on
spherical harmonics.

\end{document}